\documentclass[aps,prl,twocolumn,groupedaddress]{revtex4}
\usepackage{epsfig}
\begin{document}

\title{The chiral Anomalous Hall effect in re-entrant ${\bf Au}$Fe  alloys}
\author{F. Wolff Fabris $^{\star}$, P.Pureur, J. Schaf }
\affiliation{Instituto de F\'{i}sica, Universidade Federal do Rio Grande do Sul, \\
Caixa Postal 15051, 91501970 Porto Alegre, RS, Brazil}
\author{ V.N. Vieira}
\affiliation{Instituto de F\'{i}sica e Matem\'{a}tica, Universidade Federal do Pelotas,\\
Caixa Postal 354, 96010900 Pelotas, RS, Brazil}
\author{I.A. Campbell}
\affiliation{Laboratoire des Collo\"{\i}des, Verres, et Nanomateriaux, Universit\'{e} Montpellier II,\\
34095 Montpellier, France}


\begin{abstract}

The Hall effect has been studied in a series of ${\bf Au}$Fe samples in the re-entrant concentration range, as well as in part of the spin glass range. An anomalous Hall contribution linked to the tilting of the local spins can be identified, confirming theoretical predictions of a novel topological Hall term induced when chirality is present. This effect can be understood in terms of Aharonov-Bohm-like intrinsic current loops arising from successive scatterings by canted local spins. The experimental measurements indicate that the chiral signal persists, meaning scattering within the nanoscopic loops remains coherent, up to temperatures of the order of $150 K$. 
\end{abstract}


\maketitle

\section{Introduction}
The "anomalous" ferromagnetic contribution to the Hall signal was first observed by Hall shortly after his discovery of the ordinary Hall effect, and was studied in detail by A.W. Smith in 1910 \cite{smith:10}. For many decades, the accepted parametrization of the Hall resistivity in magnetic conductors has been in terms of the canonical expression  
\begin{equation}
\rho_{xy}(T) = R_{h}(T)B_z = R_{0}(T)B_z + R_{s}(T)M_z(T)
\end{equation}
where $B_z$ is the applied field, $M_z(T)$ is the global magnetization, $R_s/\mu_0$ is the anomalous Hall effect (AHE) coefficient and $R_{0}(T)$ the ordinary (or Lorentz) Hall coefficient. Recently the Karplus-Luttinger (KL) "anomalous velocity" term \cite{karplus:54} which is usually the major contribution to the AHE in band ferromagnets has been re-interpreted in terms of the k-space Berry phase \cite{jungwirth:02,onoda:03,fang:03,yao:04,haldane:04}, giving new insight into the origin of the intrinsic AHE, and allowing explicit estimates from band structure calculations. This mechanism leads to an intrinsic Hall current and hence through the definitions of the coefficients to a term in $R_{s}(T)$ proportional to the square of the longitudinal resistivity $\rho(T)$. Otherwise the KL AHE depends only on the band structure and not on the electron scattering. 

But the KL term is not the only contribution to the AHE and extrinsic terms (skew scattering and side jump) also exist. In addition, for conductors containing spins whose local magnetic axes are tilted away from the global magnetization direction, on theoretical grounds a further AHE term has recently been predicted. This can be described as a physically distinct Berry phase contribution occuring in real space when the spin configuration is topologically nontrivial; data on regularly ordered systems such as magnetites and perovskites whose spins are tilted have been interpreted assuming a supplementary AHE contribution of this type in the analysis \cite{ye:99,taguchi:01,taguchi:03}. The presence of this term is remarkable because it involves the magnetization components perpendicular to ${\bf M}$. The theoretical principles of this contribution, intrinsically linked to chirality, have now been spelt out for the specific case of disordered systems with canted spins such as spin glasses and re-entrant ferromagnets \cite{tatara:02,kawamura:03}. The coupling between the magnetization and the spin chirality through the spin-orbit interaction leads to a non-zero net chirality when there is a finite magnetization which is either induced by a magnetic field in the spin glass case, or which is spontaneous in the re-entrant case \cite{ye:99,tatara:02,kawamura:03}. However an {\it a priori} estimate of the order of magnitude for the effect in specific cases would require complex band structure calculations. 

An enlightening physical description of this term has been given by Tatara and Kohno \cite{tatara:03}. Successive coherent scatterings of an electron by three static local moments ${\bf S_1},{\bf S_2},{\bf S_3}$ whose axes are tilted away from the overall magnetization axis lead to a spontaneous loop of current whose strength is proportional to the chiral product
${\bf S_1\cdot(S_2\wedge S_3)}$. This effect is a consequence of the noncommutativity of the $SU(2)$ spin algebra which breaks the time-reversal symmetry in the scattering sequence. When an electric field $E_x$ is applied there is an overall drift of the current loops leading to a Hall current $j_y$. This description is the perturbative  analogue of the strong coupling Berry phase mechanism \cite{ye:99}. The current loops in the Tatara-Kohno description are avatars of the familar Aharonov-Bohm (AB) current loops in mesoscopic rings 
but in the disordered alloy case the loop dimensions are determined by local moment distances and so are typically nanoscopic. Also the loops are not physically isolated but exist within a macroscopic sample. As in canonical AB physics, the spontaneous currents require scattering to be coherent, but because of the small loop sizes in the chiral case this condition is less stringent than in mesoscopic samples; coherence can be expected to persist up to much higher temperatures. 

Experimentally the canting mechanism has been invoked to explain the AHE in magnetites \cite{ye:99} and Gd \cite{baily:05}, in both of which the canting is not static but is dynamic and due to thermal magnon-like excitations. We will comment on the dynamic aspect at the end of this article. In the perovskite $Nd_2Mo_2O_7$ there is weak static canting at low temperatures; an attractive explanation of the AHE based on the canting mechanism has been given by Taguchi {\it et al} \cite{taguchi:01,taguchi:03}, but this interpretation has been contested by Yasui {\it et al} \cite{yasui:06} who state that the Hall data can be analysed satisfaactorily without a canting term.         	

We have made systematic measurements of the AHE in a series of $\bf{Au}$Fe alloys covering part of the spin glass and all of the re-entrant regions of the magnetic phase diagram. Following up \cite{pureur:04} the present data confirm that there is a major contribution to the AHE linked to the presence of chirality which persists up to $T \sim 150 K$ and which can be interpreted satisfactorily in terms of the chiral AB-related mechanism \cite{tatara:02,kawamura:03,tatara:03}.

\section{Description of the experiment and the AuFe system}

We have used standard metallurgical methods to prepare ${\bf Au}$Fe alloys with nominal Fe concentrations of $8, 12, 15, 18, 21$ and $25$ atomic $\%$. Foils were prepared by cold rolling to a thickness of about $20 \mu m$, and were cut into the standard Hall geometry. After cold rolling and cutting the samples were annealed for an hour before quenching. 
Once prepared, the samples were stored in liquid nitrogen to minimize Fe migration effects which can modify the magnetic properties, particularly close to the critical concentration.
For the Hall and resistivity measurements an ac current technique was used having a sensitivity of better than  $10^{-8} V$. 
Fields up to $3T$ could be applied in the Hall geometry at temperatures from $8K$ to room temperature. The magnetization was measured independently at the same fields and temperatures with a commercial Squid magnetometer. The moment values were obtained in low demagnetization factor geometry, and were then corrected appropriately for the Hall geometry demagnetization factor. The data reported here to illustrate the observed behavior were taken with the Zero Field Cooling protocol, in fields of $0.25 T$ or $0.5 T$. These fields were strong enough for differences between Field Cooled and Zero Field Cooled signals to be neglible except for the lowest temperature points.    

The magnetic phase diagram of the $\bf{Au}$Fe alloys was established by Coles {\it et al} \cite{coles:78} almost thirty years ago, and a wide range of measuring techniques have since been used in the study of this system \cite{coles:78,sarkissian:81,campbell:86,senoussi:88,hennion:86,mirebeau:90, hennion:95}; see \cite{campbell:92} for an overview. Up to a critical concentration of about $13\%$Fe the alloys are spin glasses with the freezing temperature $T_g$ increasing regularly with concentration $c$. Then from $13\%$Fe up to about $30\%$Fe as ferromagnetic Fe-Fe near neighbor interactions begin to dominate the alloys enter a domain which has been dubbed "re-entrant" : as the temperature is lowered one first encounters a ferromagnetic ordering temperature $T_c$ which increases rapidly with $c$ and then a second "canting" temperature $T_k$, one of whose signatures is a dramatic drop in the low field ac susceptibility. $T_k(c)$ drops regularly with increasing Fe concentration $c$. One now knows that below $T_k$ the system still has an ferromagnetic magnetization globally or within each domain; neutron depolarization proves the persistence of ferromagnetic domains down to the lowest temperatures \cite{mirebeau:90} in the re-entrant region but the individual Fe spins become statically canted locally with respect to the global or domain magnetization axis. Neutron diffraction shows that the transverse spin components in the re-entrant phase are not entirely random but that there are transverse ferromagnetic correlations between the spins \cite{hennion:86}. The drop in susceptibility is due to domain wall pinning through the onset of Dzaloshinski-Moriya interactions when canting sets in \cite{campbell:86,senoussi:88}. The usual canting temperature $T_k$ estimates correspond to measurements using static or low frequency techniques, but for temperatures between $T_c$ and $T_k$ inelastic neutron diffraction (which is a high frequency measurement) shows magnon softening indicating a slowing down of canting dynamics above $T_k$ \cite{hennion:95}. 

For present purposes this alloy series has two main advantages. First, the basic electronic structure of the alloys is that of a noble metal containing  transition metal sites and so can be considered to be relatively simple, in contrast to those of the systems in which chiral AHE effects have been invoked up to now. The resistivities are high throughout ( typically $80 \mu\Omega cm$) because of strong magnetic impurity scattering \cite{mydosh:74}. With this type of electronic structure, the KL term can be expected to be dominant and should behave rather regularly both as a function of temperature and of concentration. Assuming that the effective band structure can be concentration dependent but can be considered to remain essentially independent of temperature at each concentration, the KL transverse resistivity may be written as 
\begin{equation}
\rho_{xy}(KL) = \lambda (c)M_z(T)\rho^2(T)
\end{equation}
where $\lambda(c)$ is a concentration dependent parameter.
The temperature dependence of the KL term was discussed recently \cite{zeng:06}  in the case of the $Mn_5Ge_3$ local moment ferromagnetic compound. As the basic electronic structure of the ${\bf Au}$Fe alloys is simple (in contrast to that of the ferromagnetic perovskite $SrRuO_3$ for instance \cite{fang:03,kats:04}) one should expect that a temperature independent $\lambda (c)$ in the KL term for each alloy should be a reasonable approximation. It can be noted that below $T_c$ the absolute value of the calculated $R_{h}^{*}(T)$ tends to drop as $T$ decreases in the re-entrant alloys because the drop in resistivity more than compensates the increase of magnetization.

\begin{figure}
\includegraphics[width=7cm, height=10cm,angle=0]{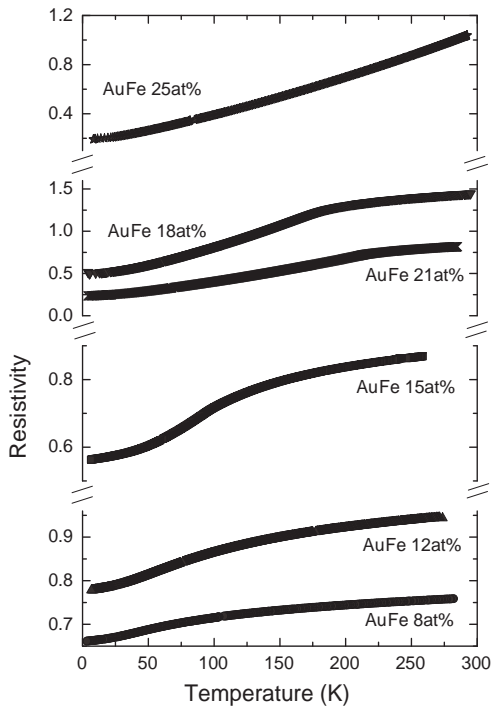}
\caption{The longitudinal resistivity in $\mu\Omega$m  of the different samples as functions of temperature}
\protect\label{fig:rho}
\end{figure}

\section{Hall measurements and analysis}

Hall data measured at lower fields for the region of  $T_g$ in the spin glass alloys ${\bf Au}$Fe and ${\bf Au}$Mn  containing $8\%$ impurity have been analysed to provide evidence for the existence of a chiral term \cite{pureur:04,taniguchi:04}, and preliminary data on the re-entrant region were given.  

Figure 1 shows the longitudinal resistivity at zero applied field of the series of ${\bf Au}$Fe samples as  functions of the temperature.

The Hall coefficent $R_{h}(T)$ is shown in Figures 2 to 6 as a function of the product   $M_{h}(T)\rho(T)^2$ at a single fixed magnetic field for each of six alloys. Here $M_{h}(T)$ is the measured ratio of the magnetization $M(T)$ in Hall geometry to the applied field, $M_{h}(T)= M(T)/B$.  For these measurements the applied fields have been chosen such that the re-entrant samples are close to  technical saturation at low $T$. If we  assume that only the canonical Lorentz and KL Hall terms contribute then we expect to observe :
\begin{equation}
R_{h}^{*}(T) = R_0(c) + \lambda (c)M_{h}(T)[\rho(T)]^2
\label{Rh_eqn}
\end{equation}
where the first term is the ordinary Hall coefficient and the second term is the KL AHE contribution. (The product $M_{h}(T)[\rho(T)]^2$ is denoted $A(T)$ in Figures 2 to 6). This relation should hold for both the paramagnetic and ferromagnetic temperature ranges, and if we assume that $R_0$ and $\lambda (c)$ are temperature independent for any given sample, the data plotted as $R_{h}(T)$ against $A(T)$ should lie on a straight line. We ignore possible skew scattering terms because in spin glass alloys these have always been found to be weak compared to the KL contributions except for concentrations much lower than those studied here. Indeed for each sample the data in the high temperature range do fall on a straight line, which is consistent with the assumption that the two conventional terms alone explain the observed behavior at high  $T$. The intercept and the slope of each line provide us with values of $R_0(c)$ and $\lambda (c)$ respectively in equation \ref{Rh_eqn}, and both $R_{0}(c)$ and $\lambda (c)$ turn out to be strongly concentration dependent.  In these concentrated alloys and at the fields indicated $R_0$ makes only  a relatively small contribution to the total $R_{h}$ except towards the very high temperature limit. For the lower concentrations its value is close to that of Au metal, $-7.10^{-11} m^3/C$ \cite{hurd:72} but $R_0(c)$ then evolves towards positive values, changing sign near 13$\%$ Fe (see \cite{mcalister:76,barnard:88}). For the concentrations for which we show data,  $\lambda(c)$ can be estimated accurately from the $R_{h}(T)$ against $M_{h}(T)[\rho(T)]^2$ plots. $\lambda (c)$ evolves steadily from negative at low Fe concentrations to positive at high concentrations with a change of sign at about $16\%$Fe. The behaviors of both $R_0(c)$ and $\lambda (c)$ as functions of concentration are very similar to those of the purely ferromagnetic  ${\bf Ni}$Fe and ${\bf Pd}$Fe alloy series which one can expect {\it a priori} to have a broadly similar electronic structures to the ${\bf Au}$Fe series.  (The AHE exponent $R_s(c)$ passes from negative to positive near $13\%$Fe in ${\bf Ni}$Fe \cite{jellinghaus:60,campbell:70} and near $18\%$Fe in ${\bf Pd}$Fe \cite{abramova:74}).  

A "conventional" AHE $R_h ^{*}(T)$ was then calculated over the entire temperature range assuming  $\lambda (c)$ and $R_0(c)$ to remain temperature independent down to low temperatures. These calculated results are shown in figs 7
to 11 as functions of the temperature together with the experimentally
determined $R_h(T)$.  The deviations of the observed $R_{h}(T)$ curve from the calculated $R_h ^{*}(T)$  is a signature of the appearance of an additional contribution to the AHE.
The $25\%$Fe sample shows only a minor negative deviation, while for each of the lower concentrations there is a striking difference between the measured $R_{h}(T)$ and the $R_{h}^{*}(T)$ curve calculated with the conventional contributions only. The total  $R_{h}(T)$ even changes sign with temperature for the intermediate concentrations. 

\begin{figure}
\includegraphics[width=6.6cm, height=5cm,angle=0]{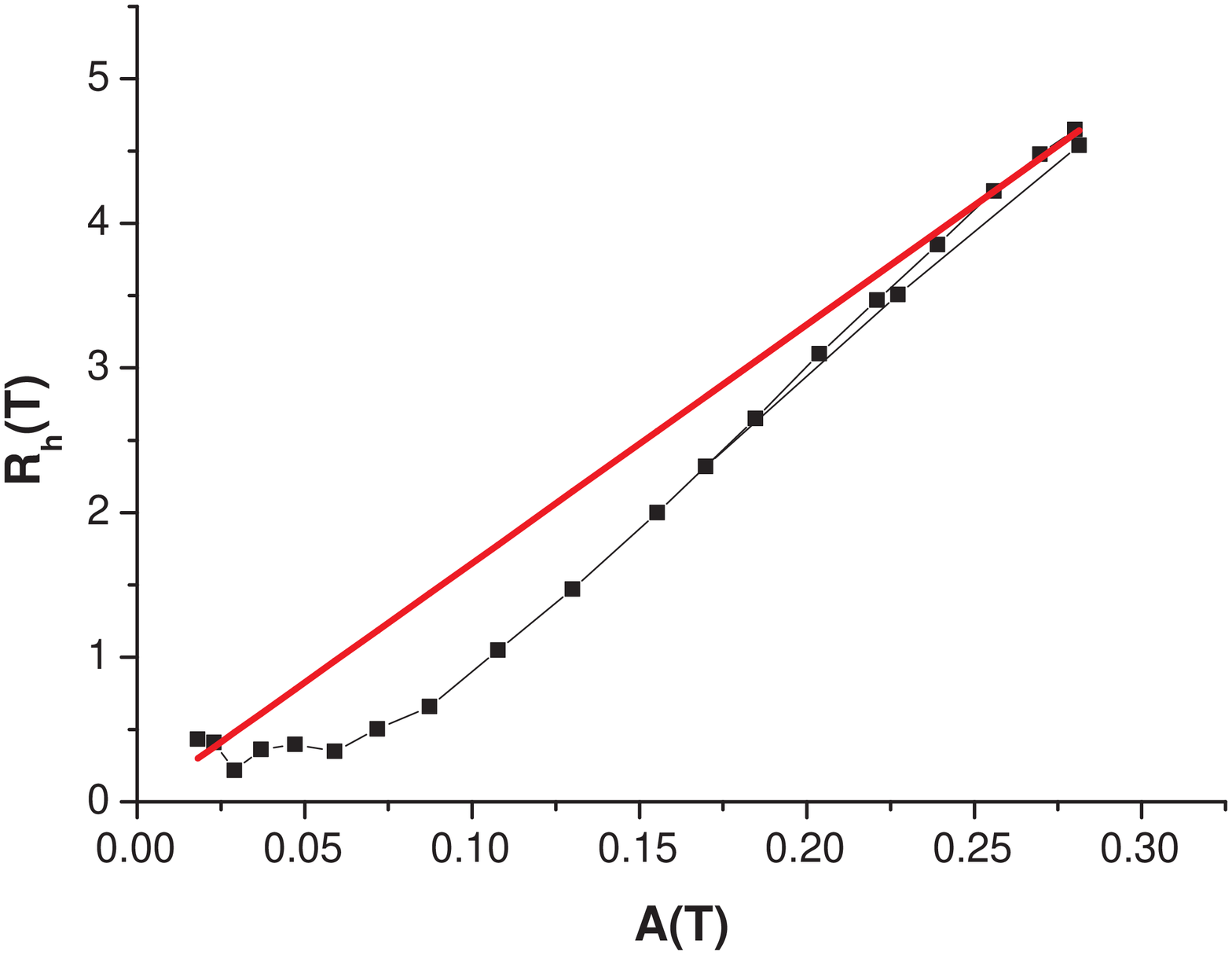}
\caption{(Color on line){\bf Au}Fe25\%. In this and following figures the total Hall coefficient $R_h(T)$ is given in units of $10^{-9}m^3/C$. The applied fields are  $0.5 T$ for the two highest concentrations, and of  $0.25T$ for the others. The "conventional" coefficient $R_{h}^{*}(T)$ is calculated for the same field assuming that only the standard ordinary Hall and KL terms contribute (see equation (2)). $R_{h}^{*}(T)$  is shown as a straight [red] line. In Figures 2 to 6 the $x$ axis is $A(T) = M(T)\rho(T)^2$. }
\protect\label{fig:AF25_RhA}
\end{figure}

\begin{figure}
\includegraphics[width=6.6cm, height=5cm,angle=0]{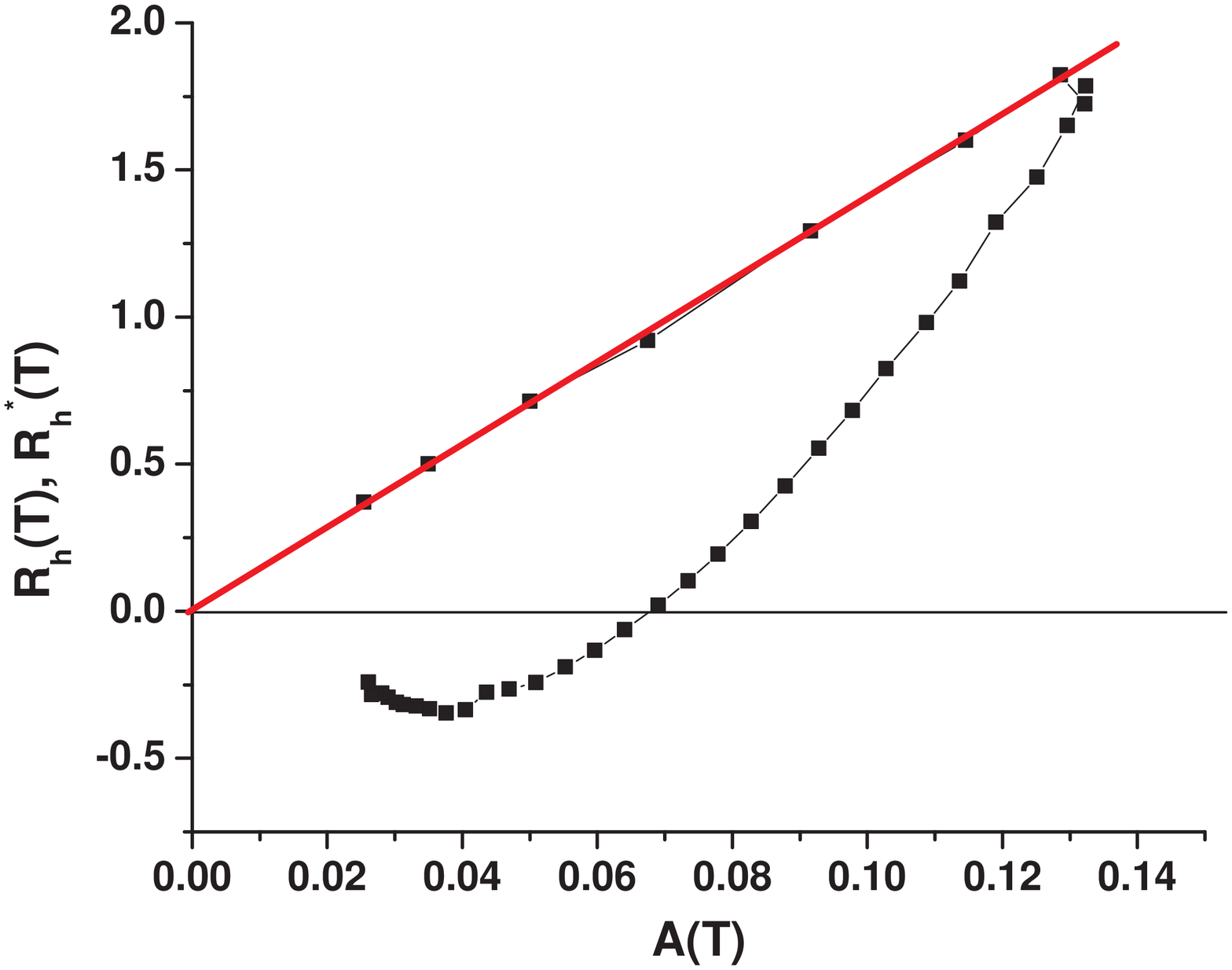}
\caption{(Color on line){\bf Au}Fe21\%. The Hall coefficient $R_h(T)$ and $R_{h}^{*}(T)$ (straight [red] line) as in Figure 2, with $A(T) = M(T)\rho(T)^2$ on the $x$ axis.}
\protect\label{fig:AF21_RhA}
\end{figure}

\begin{figure}
\includegraphics[width=6.6cm, height=5cm,angle=0]{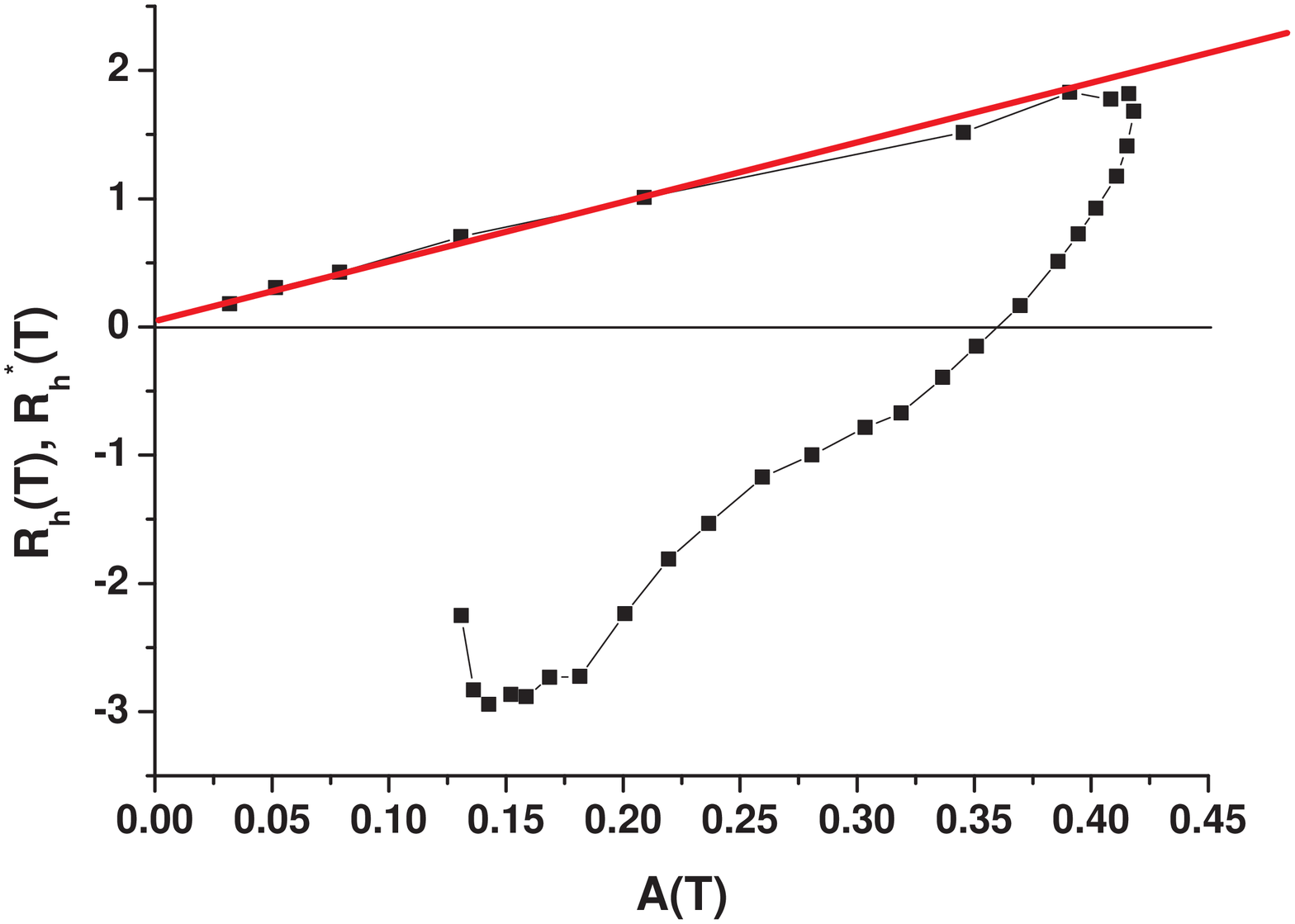}
\caption{(Color on line){\bf Au}Fe18\%. The Hall coefficient $R_h(T)$ and $R_{h}^{*}(T)$ (straight [red] line) as in Figure 2, with $A(T) = M(T)\rho(T)^2$ on the $x$ axis.}
\protect\label{fig:AF18_RhA}
\end{figure}

\begin{figure}
\includegraphics[width=6.6cm, height=5cm,angle=0]{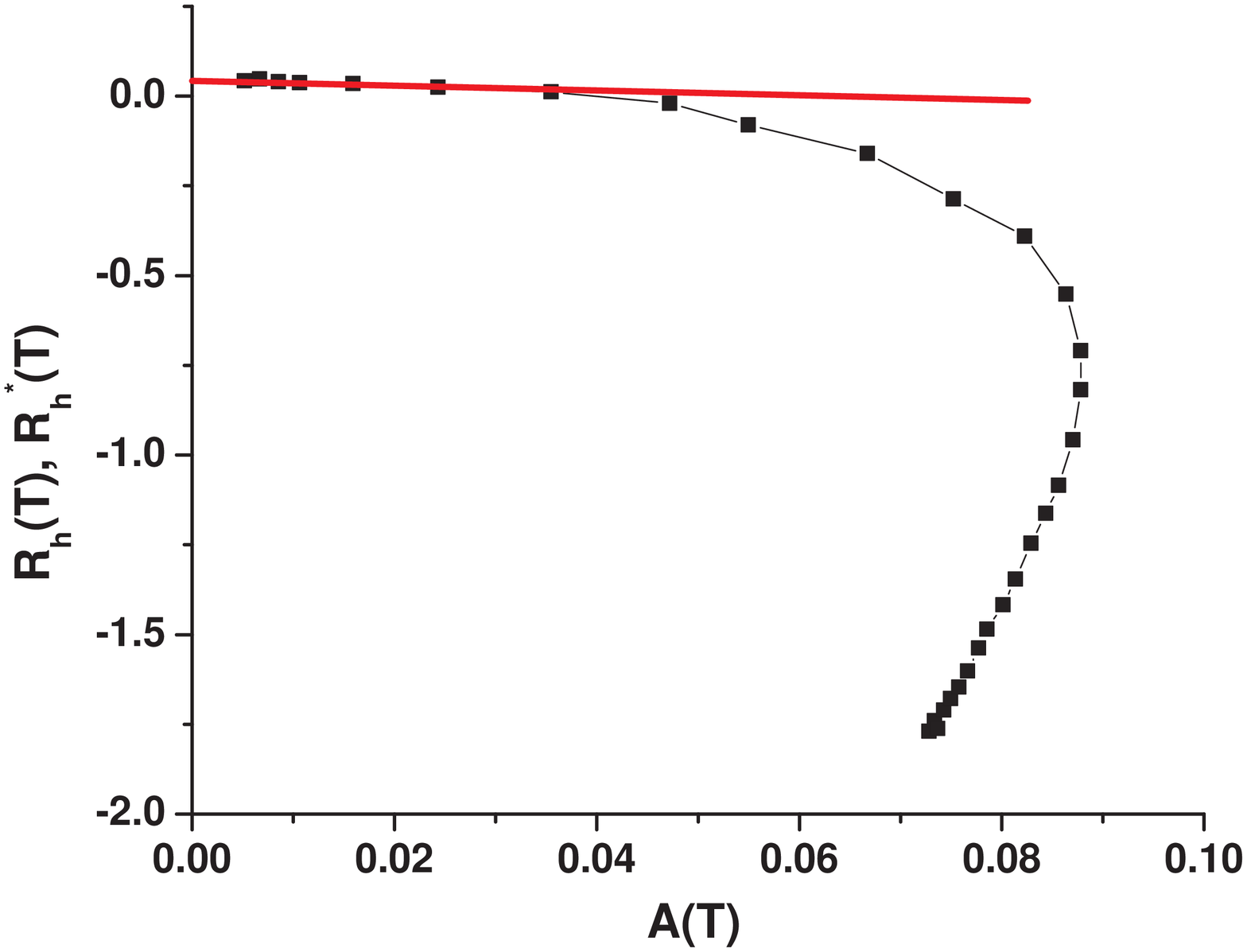}
\caption{(Color on line){\bf Au}Fe15\%. The Hall coefficient $R_h(T)$ and $R_{h}^{*}(T)$ (straight [red] line) as in Figure 2, with $A(T) = M(T)\rho(T)^2$ on the $x$ axis.}
\protect\label{fig:AF15_RhA}
\end{figure}

\begin{figure}
\includegraphics[width=6.6cm, height=5cm,angle=0]{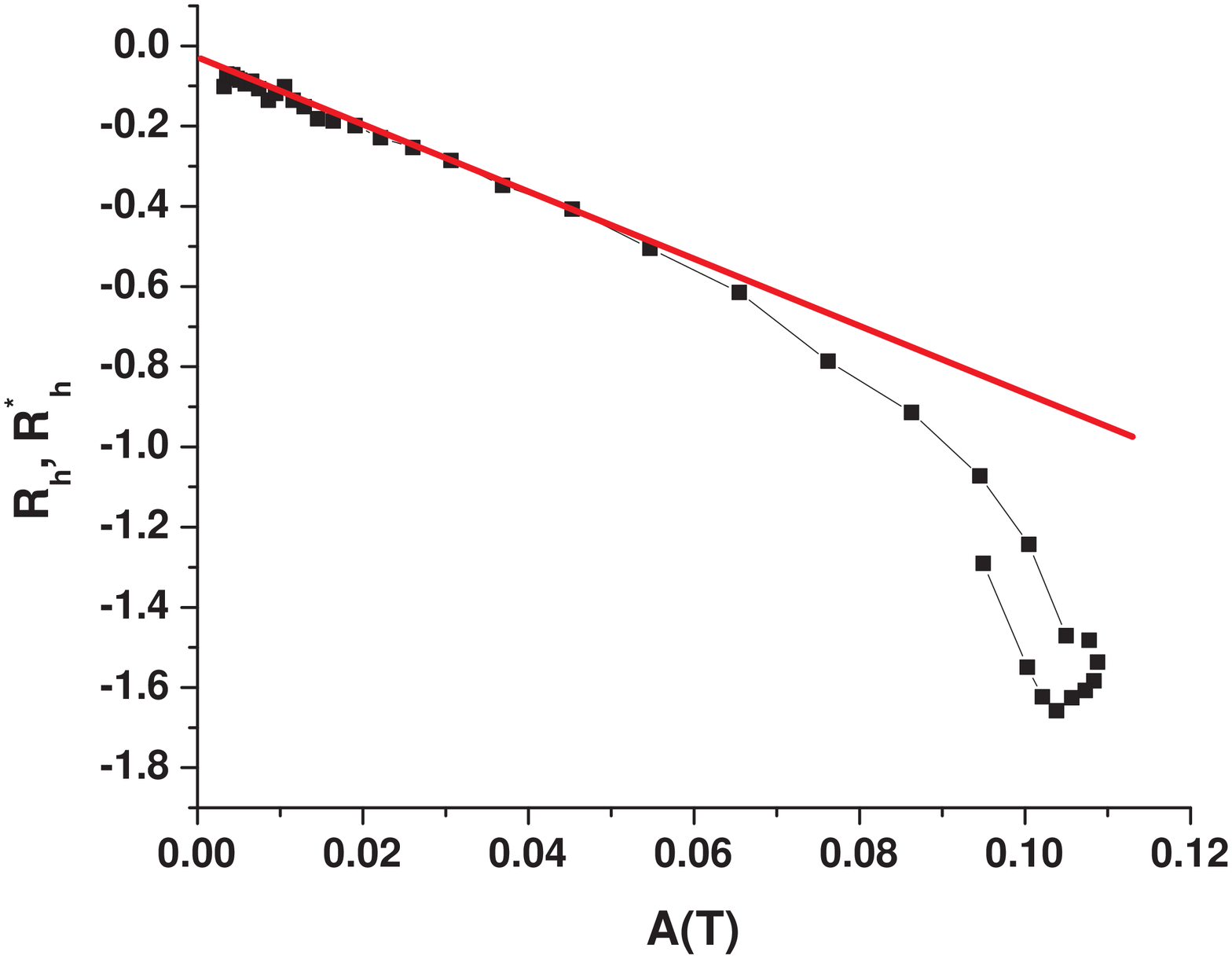}
\caption{(Color on line){\bf Au}Fe12\%. The Hall coefficient $R_h(T)$ and $R_{h}^{*}(T)$ (stright [red] line) as in Figure 2, with $A(T)= M(T)\rho(T)^2$ on the $x$ axis.}
\protect\label{fig:AF12_RhA}
\end{figure}

\begin{figure}
\includegraphics[width=6.6cm, height=5cm,angle=0]{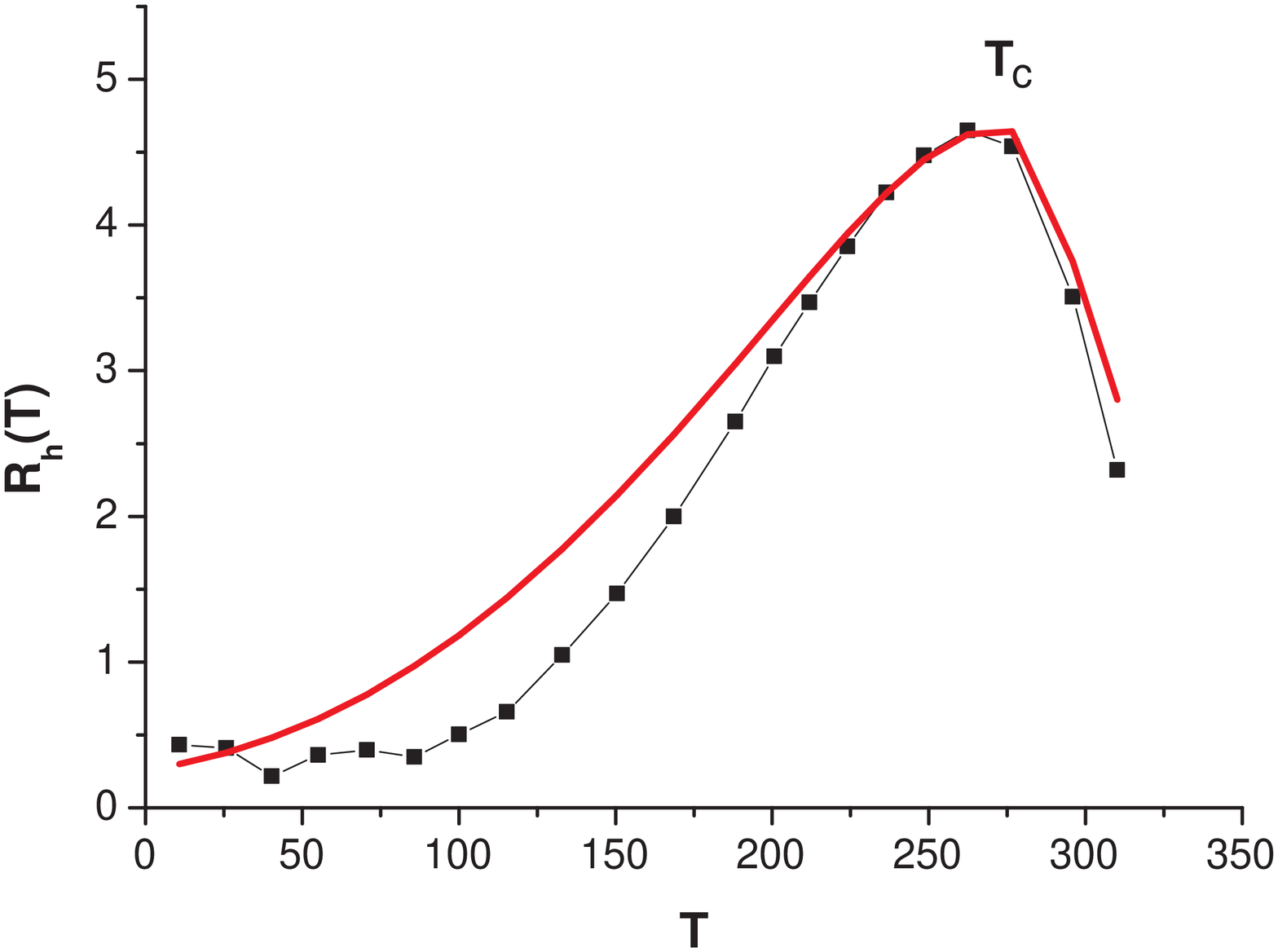}
\caption{(Color on line){\bf Au}Fe25\%. The Hall coefficient $R_h(T)$ and $R_{h}^{*}(T)$ (red curve) as in Figure 2, with temperature $T$ on the $x$ axis. The Curie temperature $T_c$ and the quasi-static canting temperature $T_k$ are indicated.}
\protect\label{fig:AF25_RhT}
\end{figure}

\begin{figure}
\includegraphics[width=6.6cm, height=5cm,angle=0]{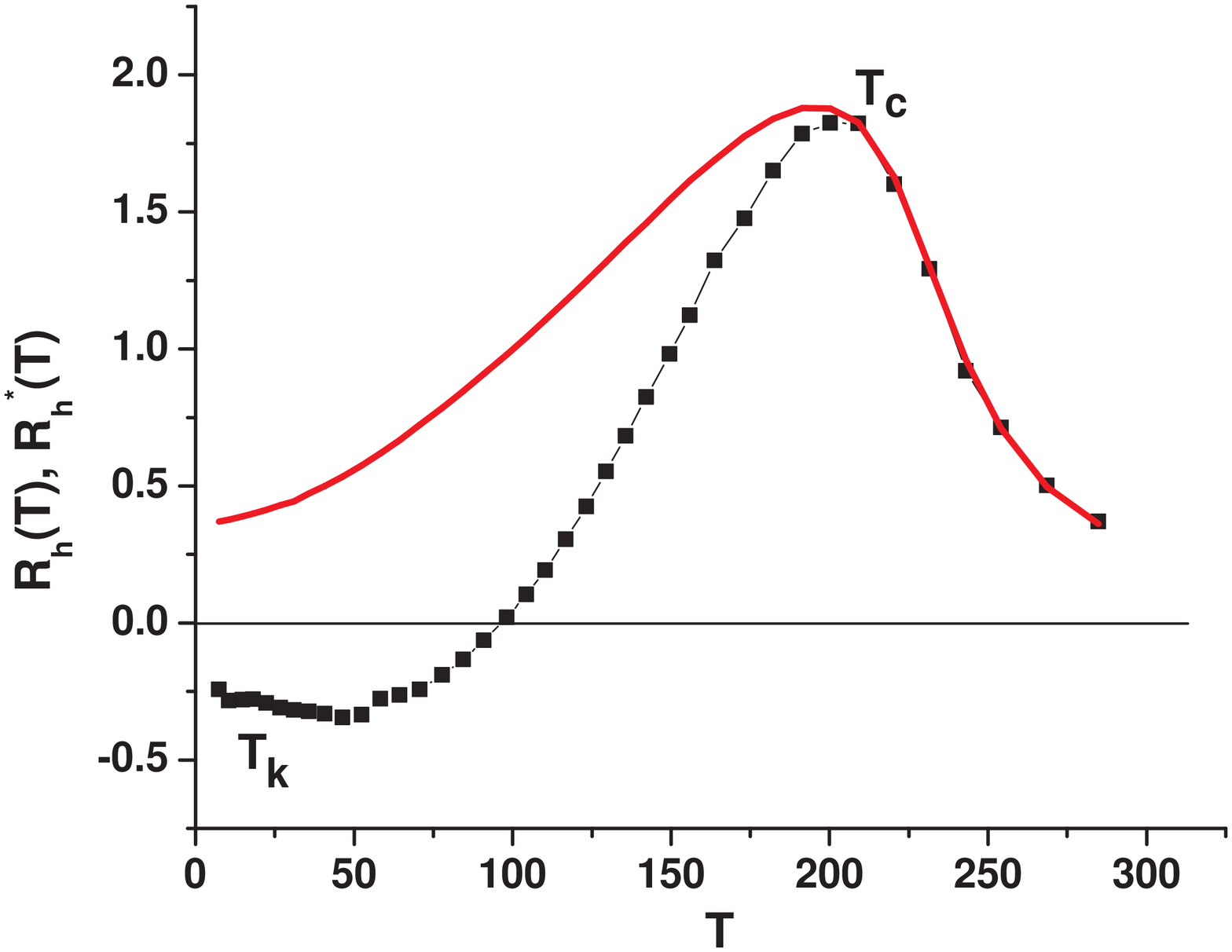}
\caption{(Color on line){\bf Au}Fe21\%. The Hall coefficient $R_h(T)$ and $R_{h}^{*}(T)$ (red curve) as in Figure 2, with temperature $T$ on the $x$ axis. The Curie temperature $T_c$ and the quasi-static canting temperature $T_k$ are indicated.}
\protect\label{fig:AF21_RhT}
\end{figure}

\begin{figure}
\includegraphics[width=6.6cm, height=5cm,angle=0]{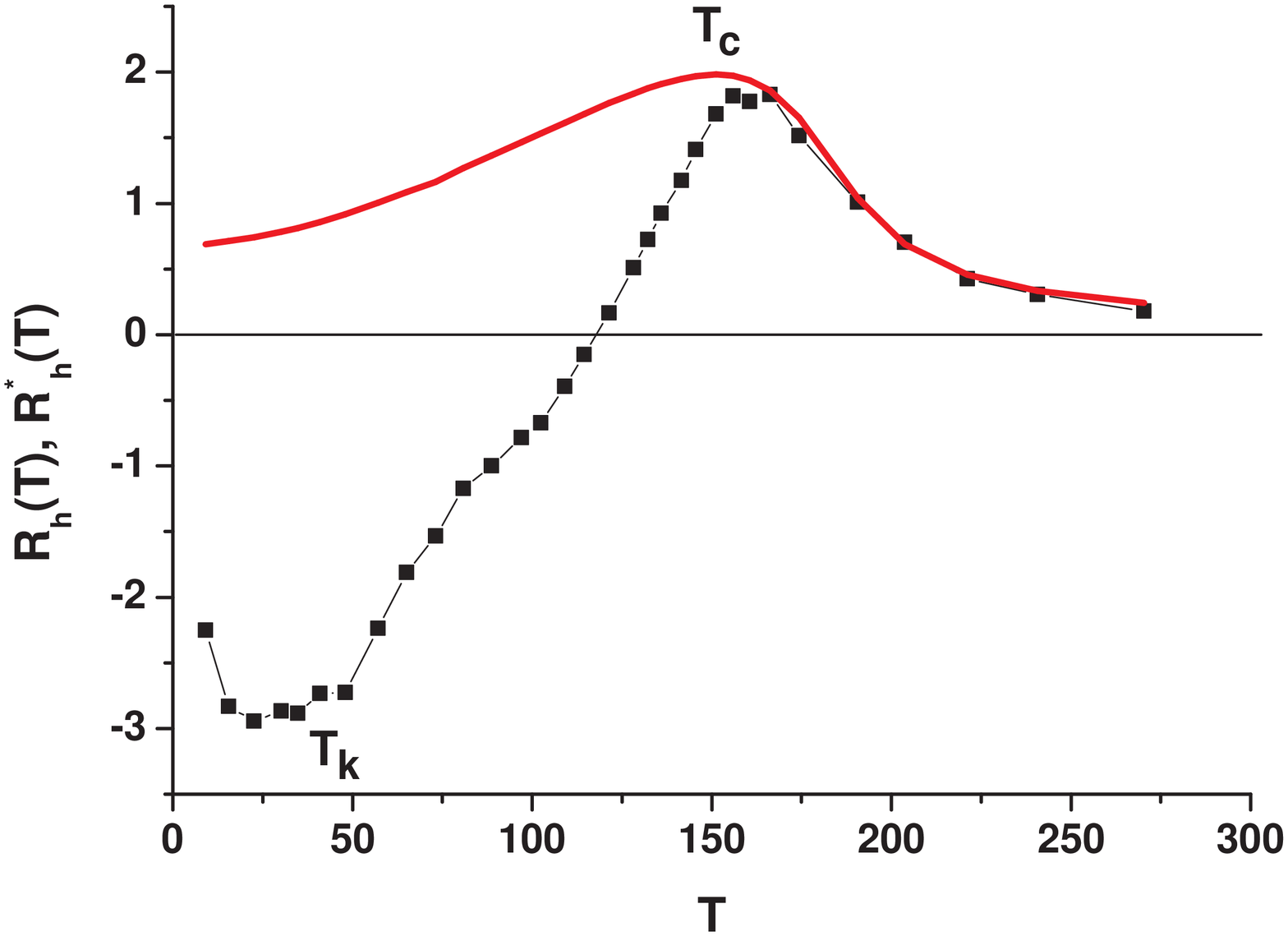}
\caption{(Color on line){\bf Au}Fe18\%. The Hall coefficient $R_h(T)$ and $R_{h}^{*}(T)$ (red curve) as in Figure 2, with temperature $T$ on the $x$ axis. The Curie temperature $T_c$ and the quasi-static canting temperature $T_k$ are indicated.}
\protect\label{fig:AF18_RhT}
\end{figure}

\begin{figure}
\includegraphics[width=6.6cm, height=5cm,angle=0]{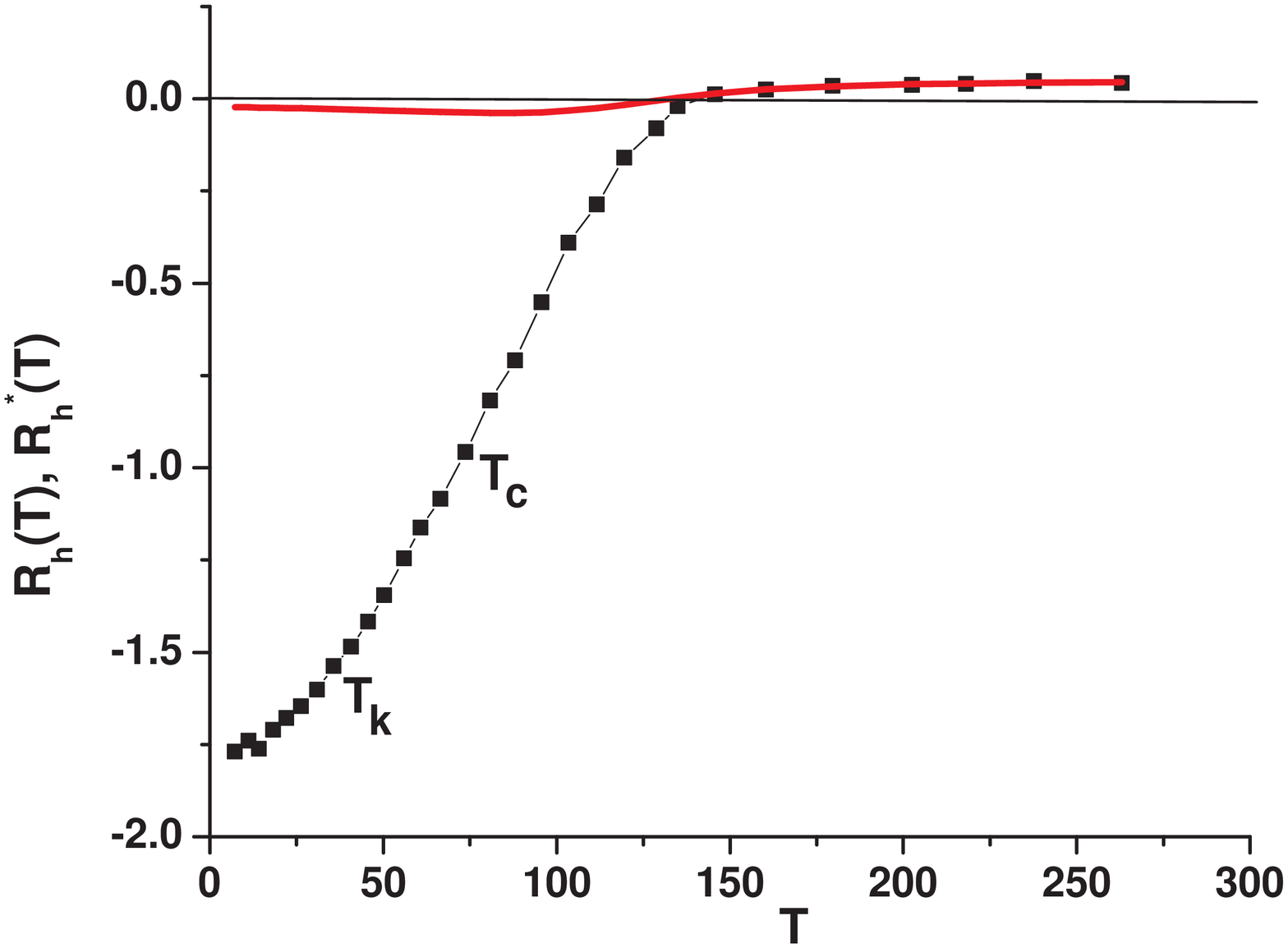}
\caption{(Color on line){\bf Au}Fe15\%. The Hall coefficient $R_h(T)$ and $R_{h}^{*}(T)$ (red curve) as in Figure 2, with temperature $T$ on the $x$ axis. The Curie temperature $T_c$ and the quasi-static canting temperature $T_k$ are indicated.}
\protect\label{fig:AF15_RhT}
\end{figure}

\begin{figure}
\includegraphics[width=6.6cm, height=5cm,angle=0]{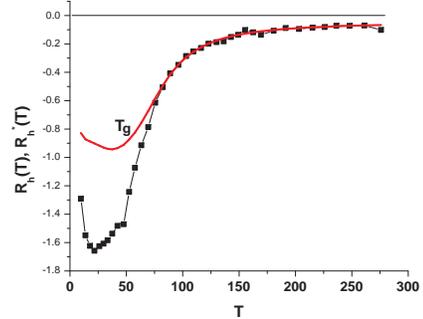}
\caption{(Color on line){\bf Au}Fe12\%. The Hall coefficient $R_h(T)$ and $R_{h}^{*}(T)$ (red curve) as in Figure 2, with temperature $T$ on the $x$ axis. The spin glass temperature $T_g$ is indicated.}
\protect\label{fig:AF12_RhT}
\end{figure}

We ascribe the difference $[R_h(T)-R_h^{*}(T)]$ to a chiral term. Consider first the high concentration end. At $25\%$Fe, where we know that low temperature canting is weak (but not strictly zero \cite{rakoto:84}), the KL term which is positive at this concentration $R_h(T)$ dominates over the whole temperature range. 
Then as the Fe concentration is lowered, we know that the low temperature canting of the Fe spins becomes progressively stronger. For these alloys, at low $T$ a negative $[R_h(0)-R_{h}^{*}(0)]$ term steadily develops for the sequence of alloys $21\%$ Fe, $18\%$ Fe, $15\%$ Fe, $12\%$Fe. For the $21\%$ Fe and $18\%$ Fe alloys the total AHE shows a change of sign with temperature (c.f. \cite{hurd:79}), while for the $15\%$ alloy where the KL term is weak ($15\%$Fe is close to the concentration where the KL $\lambda (c)$ factor changes sign) the negative term dominates over almost the entire temperature range up to about $150 K$. Finally when we pass the transition into the spin glass alloy region, for the $12\%$Fe there remains a negative contribution with respect to the calculated $R_{h}^{*}(T)$  which peaks in the neighbourhood of $40 K$. 
It is important to note that in all these alloys where the difference term $[R_h(0)-R_{h}^{*}(0)]$ can be clearly identified it is always negative, and persists up to temperatures of the order of $150K$ whatever $T_c(c)$ or $T_k(c)$ \cite{tanaguchi_comment}. 

The difference term then appears to evolve to positive by $8\%$Fe as was observed at lower applied fields \cite{pureur:04}. The change in sign in the canting term may be associated with the difference between ferromagnetic correlations among the canted spin components for the more concentrated alloys as compared to quasi-random correlations well in the spin glass region.

The present experimental data demonstrate that in the re-entrant alloys there is indeed a large negative contribution in addition to the canonical KL term, and that the strength of this contribution is closely correlated with the degree of canting. At low temperatures, this term is large enough to dominate the KL term over almost the entire re-entrant region. Because of the clear correlation with the presence of canting, the difference term can be confidently identified with the theoretically predicted chiral or real space Berry phase term \cite{tatara:02,kawamura:03}. It can be noted that in the presence of the chiral AHE the standard Equation~(1) can still be written down formally, but it loses all transparency because physical phenomena depending not only on the bulk magnetization but on the details of the transverse local spin structure and its dynamics will be hidden within the AHE parameter $R_s(T)$.

Once this point established, we can discuss the temperature dependence of the effect.  The theory \cite{kawamura:03} predicts an onset of the chiral term above as well as below the static canting temperature $T_k(c)$ because of the finite chiral susceptibility, and the data indicate inequivocally that the extra term appears already at temperatures well above $T_k(c)$ for each concentration. This can be understood at least at the phenomenological level by taking into account the relatively slow relaxation of the transverse $[x,y]$ components of the spins even above $T_k(c)$. Following the discussion of Tatara and Kohno \cite{tatara:03} the chiral Hall effect is a signature of nanoscopic spontaneous current loops due to coherent scattering by tilted spins. As the loops are small the characteristic time for an electron to undergo the series of three (or more) successive scatterings which constitute a loop is very short. As long as the scattering remains coherent and the spins remain static over this time scale (so that the canting will be sensed as frozen) the effect should persist. Thus a limiting upper temperature of the order of $150K$ is not unreasonable given the nanoscopic size of the loops, though it would be more satisfactory to have a more quantitative prediction of the expected temperature dependence of the canting term.
Further experiments to study the details of the field variation of the effect would also be of interest.   

\section{Conclusion}

The analysis of measurements of the AHE in a series of ${\bf Au}$Fe alloys demonstrates conclusively the presence of a strong AHE  a contribution linked to local spin canting in addition to the standard intrinsic KL term. The former term dominates at low temperatures over much of the concentration range, and persists up to temperatures of the order of $150K$. The results provide clear experimental evidence which supports theoretical predictions of a chiral AHE term in disordered systems possessing chiralty  \cite{ye:99,tatara:02,kawamura:03,tatara:03}. The theory shows that there should certainly be an effect, but its strength is hard to estimate. The {\bf Au}Fe alloy system turns out to be a favorable case where the canting contribution dominates, probably because of the spin-orbit interaction which is known to be strong. The chiral AHE can be understood physically in terms of a Hall current due to spontaneous nanoscopic coherent current loops \cite{tatara:03}. The Aharonov-Bohm-like current loops are a necessary consequence of time reversal symmetry breaking in sequences of three or more scatterings by tilted local spins. This mechanism has an entirely different physical origin from that of the other contributions which are invoked in interpretations of AHE data, and the present results show that it can be important even in metals with relatively simple band structures. The chiral AHE can be expected to be strongly influenced by spin dynamics through the coherence condition. 
  
The chiral Hall term should be present in any conductor containing statically canted local spins, though its relative importance will depend on factors such as the spin-orbit interaction strength. It should also appear in conductors with spins which are effectively canted thermally even if they are aligned ferromagnetically on average over long time scales, but only as long as coherence conditions are fulfilled. In particular the spin relaxation rate must be smaller than the conduction electron scattering rate. This condition does not seem to have been considered in the discussion of data on ferromagnetic systems where the AHE has been analysed in terms of the chiral effect \cite{ye:99,baily:05}.

We would like to thank Professor H. Kawamura, Dr. G. Tatara, Dr C. Pappas, Dr L. L\'evy, Dr C. Lacroix and Dr M. Taillefumier for very helpful discussions. 

$^\star$ Present address : MST-NHMFL, Los Alamos National Laboratory, Los Alamos, New Mexico 87545,
USA.

\end{document}